\documentclass[aps,prd,preprint,nofootinbib]{revtex4}
\usepackage{amsfonts}
\usepackage{mathrsfs}
\usepackage{graphicx}
\usepackage{amsmath}
\usepackage{amssymb}
\usepackage{epsfig}
\usepackage{graphicx}
\usepackage{slashed}
\usepackage{color}
\usepackage{multirow}
\usepackage{tabularx} 
\newcommand{\appropto}{\mathrel{\vcenter{
			\offinterlineskip\halign{\hfil$##$\cr
				\propto\cr\noalign{\kern2pt}\sim\cr\noalign{\kern-2pt}}}}}

\usepackage{ulem}
\usepackage{adjustbox}

\usepackage{stmaryrd}
\newcommand{\bqa}{\begin{eqnarray}}
	\newcommand{\eqa}{\end{eqnarray}}
\newcommand{\beq}{\begin{equation}}
	\newcommand{\eeq}{\end{equation}}
\allowdisplaybreaks[1]

\graphicspath{{fig/}{dia/}} \DeclareGraphicsExtensions{.eps}
\begin{document}
	
	\title{Resolving puzzle in $\Xi_c^0\to \Xi ^-e^+ \nu_e$	with $\Xi_c-\Xi_c'$ mixing spectrum within the  quark model
	}

	\author { Chao-Qiang  Geng,  Xiang-Nan Jin  and Chia-Wei Liu\footnote{chiaweiliu@ucas.ac.cn}}
	\affiliation{	
School of Fundamental Physics and Mathematical Sciences, Hangzhou Institute for Advanced Study, UCAS, Hangzhou 310024, China
	}
	\date{\today}

	\begin{abstract}
We study the ratio of  $R=2 \Gamma(\Xi_c^0 \to \Xi^- e^+ \nu _e )/3\Gamma(\Lambda_c^+ \to \Lambda e^+ \nu _e )$, which is found to be $R= 1~(0.8)$ from the exact~(broken) $SU(3)$ flavor symmetry, in sharp contrast to the average value of $R_{av}=0.59\pm 0.10 $ from the ALICE collaboration  and   lattice QCD results. We propose to use  the mixing of $\Xi_c-\Xi_c'$ to resolve the puzzle. From the mass spectra, we find that the mixing angle  is  $|\theta_c| = 0.137(5)\pi$, which suppresses $\Xi_c\to \Xi e^+ \nu _e$ about $20\%$  model-independently, resulting in   $R\approx 0.6$ with the $SU(3)$ flavor breaking effect. We explicitly demonstrate that  $R= 0.70 \pm 0.09$ from the bag model, which is also consistent with $R_{av}$. To test the mixing  mechanism, we recommend the experiments to measure the decays of  $\Xi_c \to \Xi '(1530) e^+ \nu_e$, whose branching fractions are determined  to be $ ( 4.4\sim 8.7)\times 10^{-3}$ and $(1.3\sim 2.6 )\%$ for $\Xi_c^0$ and $\Xi_c^+$, respectively, but vanish without the mixing. In addition, nonvanishing values of  ${\cal B}(\Xi_c^+\to \Xi^{\prime 0} (1530)\pi^+ )$ and ${\cal B}(\Xi_c^+\to \Sigma^{\prime +} (1385) \overline{K}^0 )$ will also be  evidences of the mixing based on  the K\"orner-Pati-Woo theorem, which are calculated as  $(3.8\sim 7.5)\times 10^{-3}$ and $(6.6\sim 13)\times 10 ^{-4}$, respectively.  We emphasize that $\theta_c$ is sizable and should be given  serious considerations in future studies on the heavy baryon systems. 

	\end{abstract}

	\maketitle
	
	\section{Introduction}

Recently, the semileptonic decays of the charmed baryons~\cite{Cheng:2021qpd} have  raised both theoretical and experimental
interests, as they provide  clean theoretical  backgrounds to examine the standard model as well as  various specific quark models~\cite{Lc,Lc2,Lc3,Lc4,Lc5,Xic,RELA,LFQM,LFQMGENG,LFQMKE,Xic2}. 
In 2021, 
the Belle and ALICE collaborations reported the branching fractions~\cite{Belle:2021crz,ALICE:2021bli},
\begin{eqnarray}\label{1}
&&{\cal B}_{\text{Belle}}(\Xi_c^0 \to \Xi^ - e^+ \nu _e)= (1.31 \pm 0.04\pm 0.07\pm 0.38) \%\,,\nonumber\\
&&{\cal B}_{\text{ALICE}}(\Xi_c^0 \to \Xi^ - e^+ \nu_e )= 
( 2.43 \pm 0.25 \pm 0.35 \pm 0.72) \% \,,
\end{eqnarray}
where the first and second uncertainties are statistical and systematic, respectively, and the third ones come from the normalizing channel of $\Xi_c^0 \to \Xi^- \pi ^+$~\cite{abs}. This decay has also been calculated by the lattice QCD~(LQCD), with the branching fraction  found to be~\cite{Lattice}
\begin{equation}\label{2}
{\cal B}_{\text{LQCD}}(\Xi_c^0 \to \Xi^ - e^+ \nu_e )= (2.38 \pm 0.44) \%\,,
\end{equation}
which is consistent with ${\cal B}_{\text{ALICE}}$ but nearly twice larger than ${\cal B}_{\text{Belle}}$. 
Collecting  Eqs.~\eqref{1} and \eqref{2}, we arrive at the average values\footnote{We adopt the same averaging method with the Particle Data Group~\cite{pdg}. 
} 
\begin{equation}\label{bav}
{\cal B}'_{av} = ( 1.85 \pm 0.28)\,,~~~
{\cal B}_{av} = (2.39 \pm 0.40)\%\,.
\end{equation} 
Here, ${\cal B}'_{av}$ is the average of (${\cal B}_{\text{Belle}}$, ${\cal B}_{\text{ALICE}}$,  ${\cal B}_{\text{LQCD}}$), whereas ${\cal B}_{\text{Belle}}$ has not been included in 
${\cal B}_{av}$ due to its tension with the others.  

On the other hand,  very recently, the BES\MakeUppercase{\romannumeral 3} collaboration has reported the branching fraction~\cite{BESIII:2022ysa}
\begin{equation}\label{ExpLC}
	\begin{aligned}
		&	{\cal B} (\Lambda_c^+  \to \Lambda e^+ \nu_e)  = ( 3.56 \pm 0.11 \pm 0.07) \%\,,
	\end{aligned}
\end{equation}
leading to the ratios
\begin{equation}\label{eq4}
\begin{aligned}
	R(\text{Belle}) = 0.33 \pm 0.10 \,,~~R(\text{ALICE}) = 0.60\pm 0.21\,,~~R(\text{LQCD}) = 0.59 \pm 0.11 \,,
\end{aligned}
\end{equation}
where
\begin{equation}\label{Rdefinition}
R(\text{method})= 
\frac{2 \tau_{\Lambda^+_c}}{3 \tau_{\Xi^0_c}}
\frac{{\cal B}_{\text{method}}(\Xi _c^0 \to \Xi^-  e^+ \nu_e )}{{\cal B}(\Lambda_c \to \Lambda e^+ \nu_e )} \,,
\end{equation}
with $\tau_{\Lambda_c^+}$ and $\tau_{\Xi_c^0}$ the baryon lifetimes. The averages of the ratios in Eq.~\eqref{eq4}  are then given as
\begin{equation}
R_{av}'
= 0.46\pm 0.07\,,~~~
R_{av}
= 0.59\pm 0.10\,,
\end{equation}
which deviate from  $R(SU(3)_F)=1 $~\cite{SU(3),SU(3)1,SU(3)0} based on the $SU(3)$ flavor~($SU(3)_F$) symmetry by $54\%$ and $41\%$, respectively.

It is  against the common believe that the $SU(3)_F$ breaking effects shall be less than $20\%$. 
For instance, the 
 meson versions of the ratios are given by~\cite{pdg}
 \begin{eqnarray}
&&\frac{ \Gamma(D_s^+  \to \phi e ^+ \nu_e)}{\Gamma(D^+  \to \overline{K}^{\ast  0 } e ^+ \nu_e) }
= 0.91 \pm 0.06\,,~~~\frac{1}{2}\frac{ \Gamma(D_s^+  \to K^0 e^+ \nu_e)}{ \Gamma(D^+  \to \pi ^0 e ^+ \nu_e) }
=0.94\pm 0.10\,,
\end{eqnarray}
which are  expected to be unity in the exact $SU(3)_F$ symmetry and lie nicely within the $20\%$ errors. Furthermore,
even if the strange quark mass is considered,  it is found that $R=0.86$ in the relativistic quark model~\cite{RELA}, and 
$R=0.98$ and  
$R=1.18\pm 0.03$  in the light-front quark model from  Refs.~\cite{LFQM} and \cite{LFQMGENG}, respectively. 
Certainly,  there must exist an additional breaking mechanism  in $\Xi_c^{ 0} \to \Xi^- e^+ \nu_e$, which is somehow negligible 
in other decays.
In this work, we point out that the mechanism is precisely the $\Xi_c - \Xi_c'$ mixing, which is   naturally absent in other decays~\cite{SU(3)1}.

This paper is organized as follows. In Sec.~\MakeUppercase{\romannumeral 2}, we extract the 
mixing angle of $\Xi_c - \Xi_c'$ from the mass spectra. In Sec.~\MakeUppercase{\romannumeral 3}, we discuss the mixing effects in $\Xi_c \to \Xi \ell ^+ \nu_\ell $ with $\Xi_c =\Xi_c^0 (\Xi_c^+)$ for  $\Xi = \Xi^-( \Xi^0)$ and   $\ell= ( e, \mu )$. 
In Sec.~\MakeUppercase{\romannumeral 4}, we examine some of the decay channels to further confirm  the mixing in the experiments. We conclude our study in Sec.~\MakeUppercase{\romannumeral 5}.

\section{Mixing angles of  $\Xi_Q - \Xi_Q'$}

If the $SU(3)_F$ symmetry is exact, the physical baryons shall have definite $SU(3)_F$ 
 representations~\cite{Mixing}.
The $\overline{{\bf 3}}$ representation suggests that the light quark pair forms a spin-0 state, while the ${\bf 6}$ representation a spin-1 one, given as 
\begin{eqnarray}\label{spinflavor}
|\Xi_Q^{ \overline{{\bf 3} } } \rangle &=& \frac{1}{\sqrt{2}}   \left( 
\uparrow\downarrow- \downarrow\uparrow
\right)  \uparrow \otimes\frac{1}{\sqrt{2}} \left(
qsQ - sqQ
\right)\,,  \nonumber\\
|\Xi_Q ^{ {\bf 6} }  \rangle &=& \frac{1}{\sqrt{6}}   \left( 
2 \uparrow\uparrow \downarrow - \downarrow\uparrow\uparrow - \uparrow \downarrow \uparrow
\right)   \otimes \frac{ 1}{ \sqrt{2} } \left(
qsQ +  sqQ
\right) \,,
\end{eqnarray}
with $q = (u,d)$ and $Q=(b,c)$. If the $SU(3)_F$ symmetry is exact, then $\Xi_Q = \Xi_Q ^{ \overline{{\bf 3} }} $ and  $\Xi_Q' = \Xi_Q ^{ {\bf 6} }  $. 

However, the symmetry is broken by the strange quark mass. As a result, 
the physical baryons are made of liner combinations of $\overline{{\bf 3}}$ and ${\bf 6}$ instead, given as 
\begin{eqnarray}\label{ph}
	|\Xi_Q  \rangle &=&  \cos\theta _Q | \Xi_Q^{ \overline{{\bf 3}}} \rangle + \exp\left( i\phi_Q\right)  \sin \theta_Q  | \Xi_Q^{  {\bf 6}}  \rangle  \,, \nonumber\\
	|\Xi_Q  ' \rangle &=&   \cos \theta_Q | \Xi_Q^ {{\bf 6}}  \rangle  - \exp\left(- i\phi_Q\right)  \sin \theta_Q | \Xi_Q^{ \overline{{\bf 3}}}  \rangle 	\,,
\end{eqnarray}
where $|\Xi_Q \rangle $ and $|\Xi_Q' \rangle$ diagonalize the QCD Hamiltonian,
$ H_{\text{QCD}} | \Xi_Q  \rangle = M_{\Xi_Q} | \Xi_Q \rangle,$ $H_{\text{QCD}} | \Xi_Q' \rangle = M_{\Xi_Q'} | \Xi_Q ' \rangle,$ 
with $M_{\Xi_Q^{(\prime)}}$  the baryon masses. 
Without lost of generality, by taking $M_{\Xi_ Q'}>M_{\Xi_ Q}$ we obtain that
$M_{\Xi_c} =  ( M_{\Xi_c^0} +M_{\Xi_c^-} )/2=  2.4691(2),$ $M_{\Xi_c'} =( M_{\Xi_c^{\prime 0 }} +M_{\Xi_c^{\prime-}} )/2 =  2.5784(4),$  $M_{\Xi_ b} = ( M_{\Xi_b^0} +M_{\Xi_b^-} ) /2 = 5.7945(4),$ $ M_{\Xi_b'} = 
M_{\Xi_b^{\prime 0 } } =  
5.93502(5)$
in units of GeV~\cite{pdg}.
Here, we use the average masses of the isospin doublets to lower the uncertainties if it is  available. 

The mass matrices of $\Xi_Q$  are given as 
\begin{equation}\label{MassMatrix}
M_Q = \left(
\begin{array}{cc}
M_Q^{ \overline{ { \bf 3 }} \overline{ { \bf 3 }} } & M_Q^{ \overline{ { \bf 3 }}{\bf 6} } \\
 M_Q^{ \overline{ { \bf 3 }}{\bf 6} } &  M_Q^{ {\bf 6} {\bf 6} }
\end{array}
\right) 
=\left(
\begin{array}{cc}
	 - \cos  2 \theta_Q  & - e^{ - i\phi_Q}   \sin 2 \theta _Q \\
	-e^{ i\phi_Q } \sin 2 \theta_Q  &   \cos  2 \theta _Q
\end{array}
\right)  M_Q^\Delta  + M_Q^0  \,,
\end{equation}
with 
\begin{equation}
M^{{\bf ij}}_Q \equiv \frac{1}{\sqrt{
\langle \Xi_Q ^ {\bf i} | \Xi_Q^ {\bf i}  \rangle 
\langle \Xi_Q ^ {\bf j} | \Xi_Q^ {\bf j}  \rangle 
}}
\langle \Xi_Q ^ {\bf i} | H_{\text{QCD}} | \Xi_Q ^ {\bf j}\rangle\,,~~~~\text{for}~{\bf i},{\bf j} = \overline{{\bf 3}}, {\bf 6}\,.
\end{equation}
The off-diagonal terms, which introduce the $\Xi_Q^{{\overline{{\bf 3}}}}- \Xi_Q^{{\bf 6}}$ transition, 
can be originated by either the hyperfine  or  long distance effect. However,
the explicit mechanism is not concerned in this letter, as we extract the baryon matrix of  Eq.~\eqref{MassMatrix} based on the physical mass spectra. 
From the time-reversal symmetry, we have $\phi_Q=0$. 
It is straightforward to show that  
\begin{eqnarray}\label{thetamaster}
&& M_Q^0 = \frac{1}{2}(M_{\Xi_ Q} + M_{\Xi_Q'}  )  \,,~~~M_Q^\Delta = \frac{1}{2}(M_{\Xi_ Q'}- M_{\Xi_Q} )   \,, \nonumber\\
&&\theta_Q =  \pm \frac{1}{2} \cos ^{-1} \left( 
\frac{M_Q^{{\bf 6}{\bf 6}} - M^0 _Q  }{M_Q^\Delta }
\right)\,. 
\end{eqnarray}
Consequently, we can obtain $|\theta_Q|$ once $M^{{\bf 6}{\bf 6}}_Q$ is known. 

To obtain $M^{{\bf 6}{\bf 6}}_c $, we utilize the  mass relation\footnote{
	The mass relations in Ref.~\cite{Jenkins:1996rr} read as 
	$(M_{\Sigma_Q^*}-M_{\Sigma_Q})-2(M_{\Xi_Q^*}-M_{\Xi_Q^{\prime}})+(M_{\Omega_Q^*}-M_{\Omega_Q}) = 0$
	without considering the $\Xi_Q-\Xi_Q'$ mixing. Thus, we  replace $M_{\Xi_Q'}$ as $M_Q^{{\bf 6}{\bf6}}$ in the relations. 
	Note that there is no need to calculate 
	 the $\Lambda_Q-\Sigma_Q$ mixings  as the isospin symmetry is well protected. 
	In addition, the mixings of $\Sigma_Q^{(\prime)} -\Sigma_Q^*$ , $\Xi_Q^{(\prime)} -\Xi_Q^*$ and $\Omega_Q' - \Omega_Q^*$  are forbidden by the angular momentum conservation. We conclude that only the $\Xi_Q- \Xi_Q'$ mixing should be considered. A similar argument  can be applied to Eq.~\eqref{eq15} also.
}~\cite{Jenkins:1996rr}
\begin{eqnarray}\label{eq14}
&&M^{{\bf 6}{\bf 6}}_c = M_{\Xi_c^ *} - \frac{1}{2} \left(
M_{\Sigma_c^*} - M_{\Sigma_c} + M_{\Omega_c^*} - M_{\Omega_c} 
\right) \pm 0.46~\text{MeV} \,,
\end{eqnarray}
where 
the asterisks denote the low-lying baryons with $J=3/2$, and $\pm 0.46$~MeV correspond to the expected errors. 
On the other hand, the mass of $\Omega^*_b$ is not available yet in the experiment. Thus, we  use the improved equal spacing rule  
\begin{equation}\label{eq15}
M^{{\bf 6}{\bf 6}}_b = \frac{1}{2}\left(
M_{\Sigma_b} + M_{\Omega_b} 
\right) - \frac{1}{2} 
\left(
M_{\Sigma^\ast_c} - 2M_{\Xi_c^*}    +  M_{\Omega_c^*}  
\right) \pm 0.6~\text{MeV}
\,,
\end{equation}
from the heavy quark symmetry to fix $M_b^{{\bf 6}{\bf 6}}$ instead~\cite{Savage:1995dw,Jenkins:1996rr}. 
The errors of Eqs.~\eqref{eq14} and \eqref{eq15} are  proportional to $m_s^2/(m_QN^2_c)$ with  $m_{s,Q}$ the quark masses  and $N_c$ the color number.
We note that the same mass relations can also be obtained in the chiral effective theory, which groups the parity-odd and even baryons as chiral multiplets~\cite{Kawakami:2018olq}.

With the baryon masses from the experiments~\cite{pdg},
we arrive at 
\begin{eqnarray}\label{M66}
&&M_c^{{\bf 6}{\bf 6}} =  2.5600(11)~\text{GeV}\,, ~~~ M_b^{{\bf 6}{\bf 6}} =  5.9315(18)~\text{GeV}\,, 
\end{eqnarray}
respectively.
Comparing to $M_{\Xi_Q'}$,  it is clear that the mixings are required.
Plugging Eq.~\eqref{M66} in Eq.~\eqref{thetamaster}\,, we find that
\begin{eqnarray} \label{mixingangle}
&&|\theta_c| =   0.137 (5 )  \pi \,, ~~~|\theta_b| =     0.049 (13)  \pi\,. 
\end{eqnarray}
It indicates that about $20\%$ of $\Xi_c$ and $\Xi_c'$ are made of $\Xi_c^{{\bf 6}}$  and 
$\Xi_c^{\overline{{\bf 3}}}$, and  $3\%$ of $\Xi_b$ and $\Xi_b'$  of  
$\Xi_b^{{\bf 6}}$  and 
$\Xi_b^{\overline{{\bf 3}}}$, respectively. The ratio of $\theta_Q$ is consistent with the expectation from the heavy quark expansion, which states that $\theta_{b}/\theta_c \propto m_c/m_{b}$~\cite{Savage:1995dw}. 
In the following, we will concentrate on the charm baryon decays.


For simplicity, we take the spatial wave functions~(quark distributions) of $\Xi_c^{\overline{\bf 3}}$ and $\Xi_c ^{{\bf 6}}$ to be the same in this work, so the only difference between the two states is the spin-flavor part as shown in Eq.~\eqref{spinflavor}. 
We start with the helicity amplitudes $H_{\lambda_1 \lambda_W}$ with $\lambda_1$ and $\lambda_W$ the helicities of $\Xi$ and  off-shell $W^*$ boson, given as 
\begin{eqnarray}\label{15}
	H_{\pm \frac{1}{2}\pm  1} = \varepsilon^{*\mu}_\pm\left( N_{\text{flip}}^{eff} {\cal V}_\mu^{\uparrow \downarrow} \mp  N_{\text{flip}}^{eff} {\cal A}_{\mu}^{\uparrow \downarrow} \right) \,, \quad
	H_{\pm \frac{1}{2} z }= \varepsilon^{*\mu}_z \left( N_{\text{unflip}}^{eff}  {\cal V}_\mu^{\uparrow \uparrow} \mp N_{\text{flip}}^{eff}   {\cal A}_{\mu}^{\uparrow \uparrow} \right)  \,,
\end{eqnarray}
respectively,
where $\varepsilon^\mu$ is the polarization vector of  $W^*$~\cite{Timereversal}, 
$z\in\{ 0,t \}$, $N^{eff}$ are the  spin-flavor overlappings, given by
\begin{eqnarray}\label{19}
	N_{\text{flip}} ^{eff} \equiv 
	\left(  N_{\text{flip}}^{\overline{\bf 3}} \cos\theta_c 
	+ N_{\text{flip}}^{\bf 6}  \sin \theta_c
	\right)\,,\quad
	N_{\text{unflip}} ^{eff} \equiv
	\left(  N_{\text{unflip}}^{\overline{\bf 3}} \cos\theta_c 
	+ N_{\text{unflip}}^{\bf 6}  \sin \theta_c
	\right)\,,
\end{eqnarray}
and ${\cal V}_{\mu }^{\lambda_1 \lambda_2} $ and ${\cal A}_\mu ^{\lambda_1\lambda_2}$ are the matrix elements of the current operators
\begin{eqnarray}\label{spatio}
	{\cal V}_\mu^{\lambda_1 \lambda_2 } & \equiv &
	\left\langle q ss, J_z^3= \lambda_1 , \vec{p}=|\vec{p}| \hat{z} \right|  \overline{s}_3 \gamma_\mu c_3 \left| qsc, J_z^3 = \lambda_2 \right \rangle \,, \nonumber\\
	{\cal A}_{\mu}^{\lambda_1 \lambda_2 } &\equiv& 
	\left\langle qss, J^3_z= \lambda_1,   \vec{p}=|\vec{p}|\hat{z} \right|  \overline{s}_3 \gamma_\mu \gamma_5  c_3 \left| qsc, J^3_z = \lambda_2 \right \rangle \,,
\end{eqnarray}
with $\vec{p}$  the three-momentum of $\Xi$ in the 
rest frame of $\Xi _c$. 
Here,
the subscripts in $\overline{s}_3$ and $c_3$ indicate that they only act on the third quarks,  $J^3$ stands for the angular momentum of  the third quarks, and the states are normalized as 
\begin{equation}
\langle \Xi_{(Q)} , \vec{p}\, |\Xi_{(Q)} , \vec{p}\,' \rangle  = 2 
u^\dagger_{(Q)} u_{(Q)}
(2\pi )^3  \delta^3 (\vec{p} -\vec{p}\,' )\,,
\end{equation}
with $u_{(Q)}$ the Dirac spinor of $\Xi_{(Q)}$. 
In turn, $N^{{\bf i}}_{\text{(un)flip}}$ are defined as 
\begin{eqnarray}\label{const}
	&&N^{\bf i}_{\text{flip}} = \Big
	\langle \Xi  ,J_z =\frac{1}{2} \Big | s^\dagger \sigma_z  c \Big| \Xi ^ {\bf i}_c ,J_z =\frac{1}{2} \Big \rangle \,,~~N_{\text{unflip}}  ^ {\bf i} = 
	\Big \langle \Xi  ,J_z =\frac{1}{2} \Big | s^\dagger   c \Big| \Xi _c ^ {\bf i} ,J_z =\frac{1}{2} \Big  \rangle\,,\nonumber \\
	& & s^\dagger \sigma_z c = s^\dagger_\uparrow c_\uparrow - s^\dagger_\downarrow c_\downarrow \,,  \qquad\qquad\qquad\qquad\quad
	s^\dagger c  = s^\dagger_\uparrow c_\uparrow +  s^\dagger_\downarrow c_\downarrow \,, 
\end{eqnarray}
where
the baryon states in Eq.~\eqref{const} are in the nonrelativistic constituent quark limit. 
The values of $N^{{\bf i}}_{\text{(un)flip}}$ can be calculated once the spin-flavor parts of the wave functions are given~(see Appendix A of Ref.~\cite{Cheng:2018hwl} for instance), which are the same for all  quark models up to an overall constant.  
They are collected in TABLE~\ref{tableNflip}, where we also include the spin-flavor overlapping of $\Lambda^+_c \to \Lambda$.  


\begin{table}
	\caption{The spin-flavor overlappings defined in Eq.~\eqref{const}. }\label{tableNflip}
	\vskip 0.2cm
	\begin{tabular}{c|ccccccc}
		\hline
		\hline
		& $ \Xi_c^{\overline{ {\bf 3}}}\to \Xi  $& $ \Xi_c^{{\bf 6}} \to \Xi  $ 	&$\Lambda_c \to \Lambda $\\
		\hline
		$N_{\text{unflip}}$&$\sqrt{ \frac{3}{2} } $&$-\frac{1}{\sqrt{2}}$ &1\\
		$N_{\text{flip}}$ &$\sqrt{ \frac{3}{2} } $&$\frac{1}{3\sqrt{2}}$ &1\\
		\hline
		\hline		
	\end{tabular}
\end{table}

Finally,
the total decay widths are given as 
\begin{eqnarray}\label{number4}
&&\Gamma 
= \frac{G_F^2 }{24 \pi^3}  |V_{cs}| ^2  \int^{(M_{\Xi_c} -M_\Xi )^2}_{M_\ell^2}  \frac{(q^2-M_{\ell}^2)^2|\vec{p}|  }{8M_{\Xi_c  }^2q^2} \Big[
(\delta_\ell +1 )   \\
&&
\times \left( H_{ \frac{1}{2} 1 }^2 + H_{ -\frac{1}{2} -1 }^2   + H_{ \frac{1}{2} 0 }^2+  H_{ - \frac{1}{2} 0 }^2 \right)
 + 3 \delta_\ell \left( H_{\frac{1}{2}t} ^2 +  H_{-\frac{1}{2}t} ^2 \right)
\Big]
dq^2 \appropto (N_{\text{unflip}}^{eff} )^2 + (N_{\text{flip}}^{eff} )^2   \,,\nonumber
\end{eqnarray}
where $G_F$ is the Fermi constant, $V_{cs}=0.973$ is the Cabibbo–Kobayashi–Maskawa matrix element~\cite{pdg}, $M_{\ell}$ 
is the charged lepton mass, and $\delta_\ell = M_\ell^2 / 2q^2$.  From TABLE~\ref{tableNflip}, it is straightforward to show that $R(SU(3)_F)=1$.

We note that the sign of $\theta_c$ has a little effect on the branching fractions. It is due to that
 $N_{\text{unflip}}^{\overline{{\bf 3}}}$ and $N_{\text{flip}}^{\overline{{\bf 3}}}$ are much larger than 
  $N_{\text{unflip}}^{{\bf 6}}$ and $N_{\text{flip}}^{{\bf 6}}$. 
In addition,
 $N_{\text{unflip}}^{{\bf 6}}$ and $N_{\text{flip}}^{{\bf 6}}$ are opposite in sign, which destructively (constructively) and  constructively (destructively)  interfere with  
$N_{\text{unflip}}^{\overline{{\bf 3}}}$  and $N_{\text{flip}}^{\overline{{\bf 3}}}$ for a positive (negative)  $\theta_c$. Consequently, large parts of their effects get canceled  in the branching fractions. 
In practice, we have  the first order approximation
\begin{equation}
		{\cal B}\left( \Xi_c \to \Xi \ell^+ \nu _\ell\right) \approx  (1 - \sin ^2\theta_c ) 
		{\cal B}\left( \Xi_c^{\overline{3}} \to \Xi \ell^+ \nu _\ell \right)\,,
\end{equation}
where the second term is understood as the mixing effect. With Eq.~\eqref{mixingangle}, we find that $\Xi_c\to \Xi \ell^+ \nu$ are suppressed about $20\%$ by the mixing, which is a generic model-independent result. 
Subsequently, the $SU(3)_F$ relation shall be modified as 
\begin{equation}
R(SU(3)_F)  \to ( 1 - \sin^2 \theta_c)(1  -  0.2 ) \,,
\end{equation}
where the second term take account the $20\%$ breaking from the strange quark mass. As a result,  we obtain $R\approx 0.6$ in  agreement with $R_{av}$. Alternatively, by taking $R_{av}$ as an input, we find 
 $ |\theta_c| = (0.17 \pm 0.05)\pi$, which agrees well with   Eq.~\eqref{mixingangle}. 
 Note that $R_{av}'$ leads to   $ |\theta_c| = (0.24 \pm 0.04)\pi$, which indicates that $\Xi_c^{\overline{{\bf 3}}}$ and $\Xi_c^{{\bf 6}}$ share   $\Xi_c^{(\prime)}$ equally. 

\section{ Numerical results and discussions}

To illustrate the decays numerically,  we adopt the bag wave functions in Ref.~\cite{Liu:2022pdk,anatomy} to calculate 
 ${\cal V}^{\lambda_1\lambda_2}_\mu$ and ${\cal A}^{\lambda_1\lambda_2}_\mu$.
To explain the large $SU(3)$ flavor violation, it is important  that a theory should be able to  explain  both ${\Gamma }(\Lambda_c^+ \to \Lambda e^+\nu _e)$ and ${\Gamma }(\Xi_c^0 \to \Xi^- e^+\nu_e )$, simultaneously.  
The formalism for $\Lambda_c^+ \to \Lambda \ell^+ \nu_\ell  $ can be obtained directly by  substituting the spectator quark of $s$ as $u$ in Eq.~\eqref{spatio}. The branching fractions are found to be~\cite{anatomy}
\begin{equation}
	\begin{aligned}
				{\cal B} (\Lambda_c^+  \to \Lambda e^+ \nu_e)  = ( 3.78\pm 0.24\pm 0.05) \%\,,\\ 
		{\cal B} (\Lambda_c^+  \to \Lambda \mu ^+ \nu_\mu )  = ( 3.67\pm 0.23\pm 0.05 ) \%\,,
	\end{aligned}
\end{equation}
where the first and second uncertainties arise from the model parameters and $\tau_{\Lambda_c^+}$, respectivley. 
Notably, the results are consistent with the experimental values  in Eq.~\eqref{ExpLC}
and  LQCD calculations~\cite{Lattice Lambda}, given by
\begin{equation}
	\begin{aligned} 
&{\cal B}_{\text{LQCD}} (\Lambda_c^+  \to \Lambda e^+ \nu_e)  = ( 3.80 \pm 0.19  \pm 0.11) \%\,,\\
&{\cal B}_{\text{LQCD}} (\Lambda_c^+  \to \Lambda \mu ^+ \nu_\mu )  = ( 3.69\pm 0.19  \pm 0.11) \%	\,,
	\end{aligned} 
\end{equation}
where the uncertainties arise from the lattice simulation and $\tau_{\Lambda_c^+}$, respectively. 

\begin{figure}
	\includegraphics[width=0.45\linewidth]{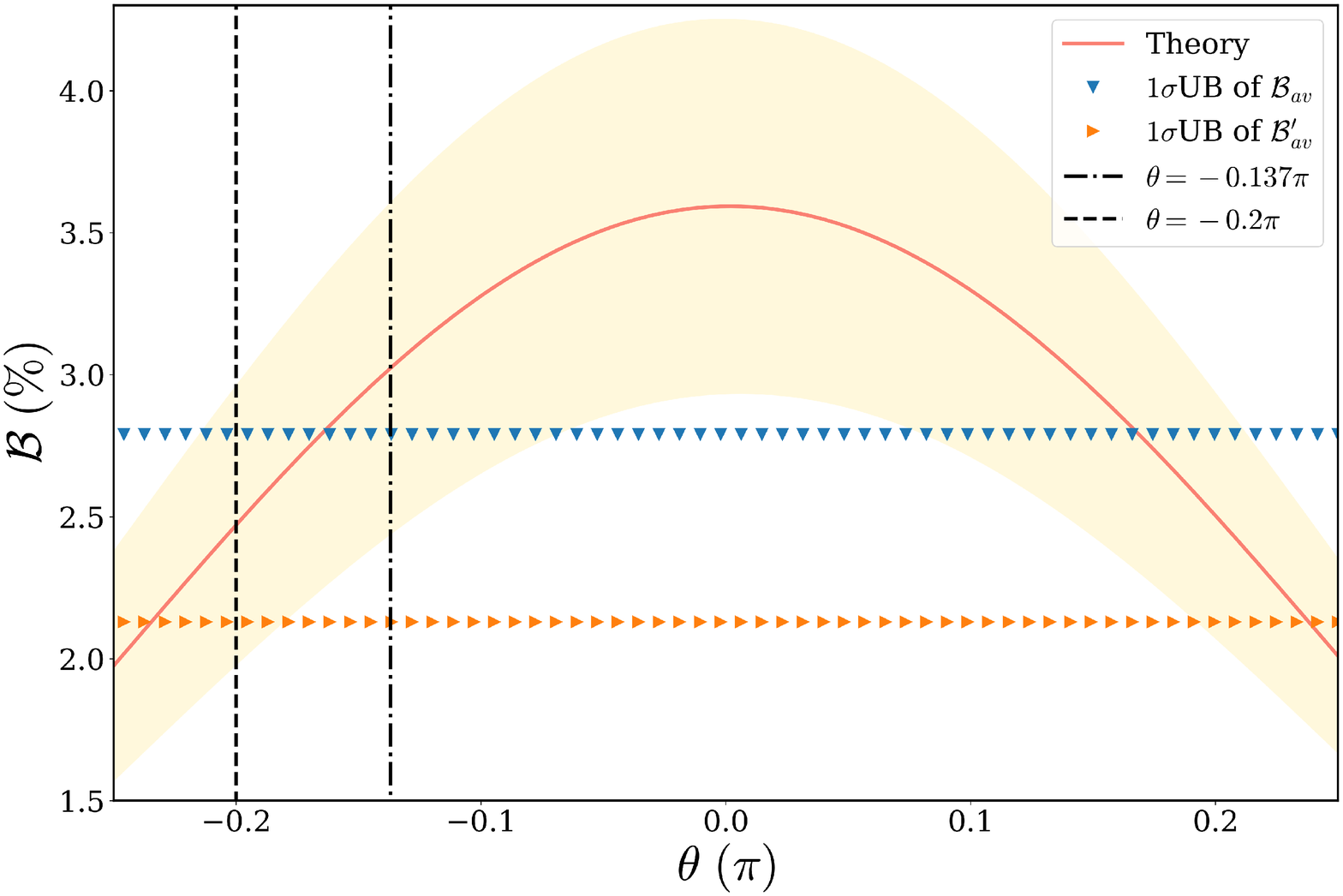}~  	
	\includegraphics[width=0.46 \linewidth]{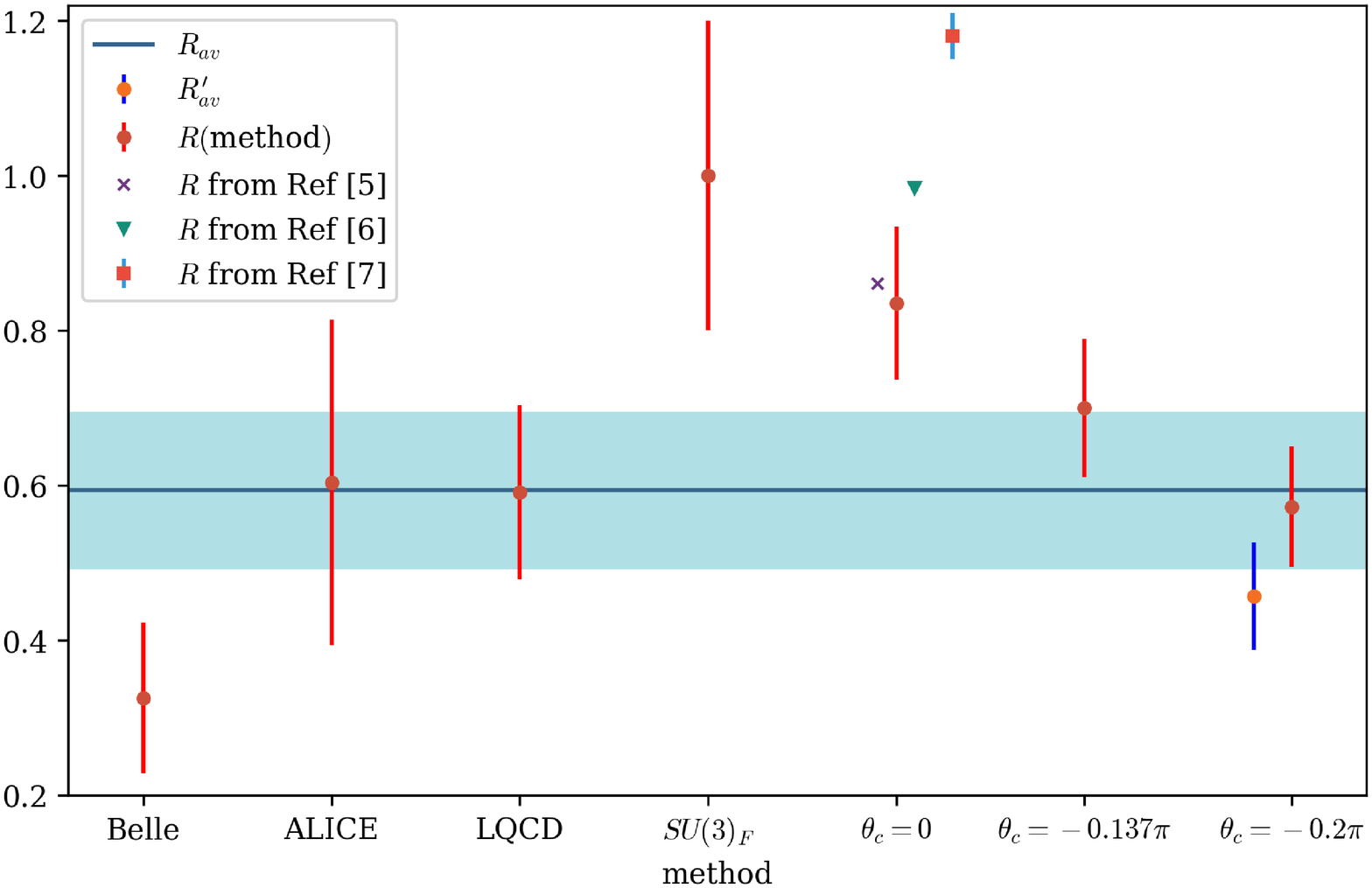}  	
	\caption{ 
		${\cal B}( \Xi_c^0 \to \Xi^- e^+ \nu_e)$ versus $\theta_c$~(left), and $R$ of different methods~(right) with the bands representing the uncertainties.
 }
	\label{PA3} 
\end{figure}

We 
now turn our attention to the $\Xi_c$ decays.  
 In the left figure of  FIG.~\ref{PA3}, we plot
${\cal B}(\Xi_c^0 \to \Xi^- e^+ \nu_e)$  versus $\theta_c$, 
where the upper bounds~(UBs) of ${\cal B}_{av}$ and ${\cal B}_{av}'$ are included for comparison. 
On the other hand,
the values with $\theta_c\in \{ \pm \pi/4, \pm 0.2 \pi , \pm 0.137\pi, 0\}$ are given explicitly in TABLE~\ref{table1}.
With $|\theta_c| = 0.137 \pi $ and $|\theta_c| = 0.2 \pi$, our predictions of the branching fractions are consistent with ${\cal B}_{av}$ within $1\sigma$. In particular,  ${\cal B}_{av}'$ can be explained by  $\theta_c =0.2\pi$. 
In contrast,  the branching fraction for $\theta_c=0$ is incompatible with  both ${\cal B}_{av}$ and ${\cal B}_{av}'$.

\begin{table}
	\caption{
The branching fractions with different $\theta_c$. 
	}\label{table1}
	\vskip 0.2cm
	\begin{tabular}{c|ccccccc}
		\hline
		\hline
$\theta_C$ &$- \pi /4 $&$-0.2\pi$ &$- 0.137 \pi $&$ 0 $&$ 0.137\pi  $ &$ 0.2\pi$ &$ \pi/4$ \\
		\hline
	$\Xi_c ^0 \to \Xi^-  e^+  \nu_e $ &$1.98(40) $&$2.47(49)$&$3.02(58)$&$3.59(66)$&$3.05(53)$&$2.50(43)$&$2.01(34)$\\
	$\Xi_c ^0 \to \Xi^-  \mu ^+  \nu_\mu  $ &$1.92 (38) $&$2.40(47)$&$2.94( 56)$&$3.49(63)$&$2.93( 51)$&$2.43(41)$&$1.95(32)$\\	
\hline
$\Xi_c^+ \to \Xi ^0  e^+  \nu_e $ &$5.89(120)$&$7.36(146)$&$9.00(174)$ &$10.7(20)$&$9.08(159)$ &$7.46(128)$ &$5.99(102)$\\
$\Xi_c^+ \to \Xi ^0  \mu ^+  \nu_\mu  $ &$5.73(115)$&$7.16(139)$&$8.75(165)$ &$10.4(19)$&$8.81(152)$ &$7.24(122)$ &$5.81(97)$\\
		\hline
		\hline		
	\end{tabular}
\end{table}

To reduce the errors and uncertainties from the model calculation, we  plot  $R(\text{method})$ defined in Eq.~\eqref{eq4}
on the right hand side  of FIG.~\ref{PA3}, where $R(\theta_c)$ are the ones computed in this work, while   $R(SU(3)_F)=1.0\pm 0.2$  along with the values
 from Refs.~\cite{LFQMGENG,LFQM,RELA}. 
We find that the result of $R(\theta_c=0)=0.84 \pm 0.10$ is consistent with  the  theoretical expectations in the literature, such as those from 
the relativistic quark model~\cite{RELA},  light front quark model~\cite{LFQM,LFQMGENG} and 
 $SU(3)_F$, but incompatible with the experiments and LQCD. 
In contrast, $R(\theta_c = -0.137\pi )=0.70\pm 0.09$  and $R(\theta_c=-0.2\pi)=0.57\pm0.08$ agree  well 
 with $R_{av}$.  
 In addition, $R_{av}'$ can be explained by $|\theta_c| = 0.2\pi$.
 We conclude that the mixing between $\Xi_c-\Xi_c'$ is supported by not only  the mass spectra but also the  semileptonic decay experiments.

\section{Mixing effects involving decuplet baryons}

To  confirm the mixing in the  future experiments, we recommend some of the decay channels involving the decuplet baryons, 
which do not exist without the  mixing.  In this work, we take $\Xi'$ to denote $\Xi(1530)^{-,0}$ for $\Xi_c = \Xi_c^{0,+}$, respectively.

\begin{figure}
	\includegraphics[width=0.3 \linewidth]{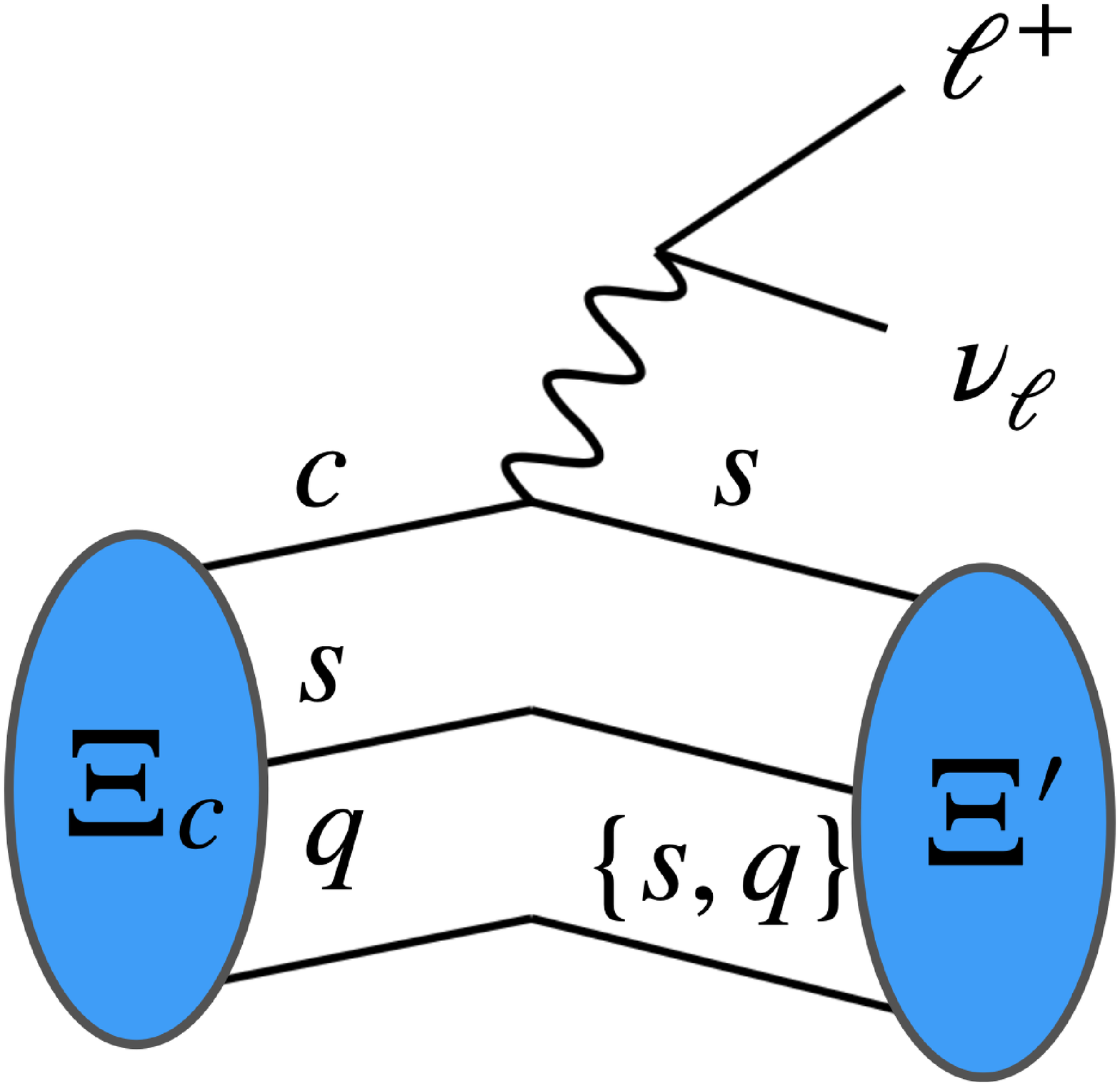}\qquad
	\includegraphics[width=0.3 \linewidth]{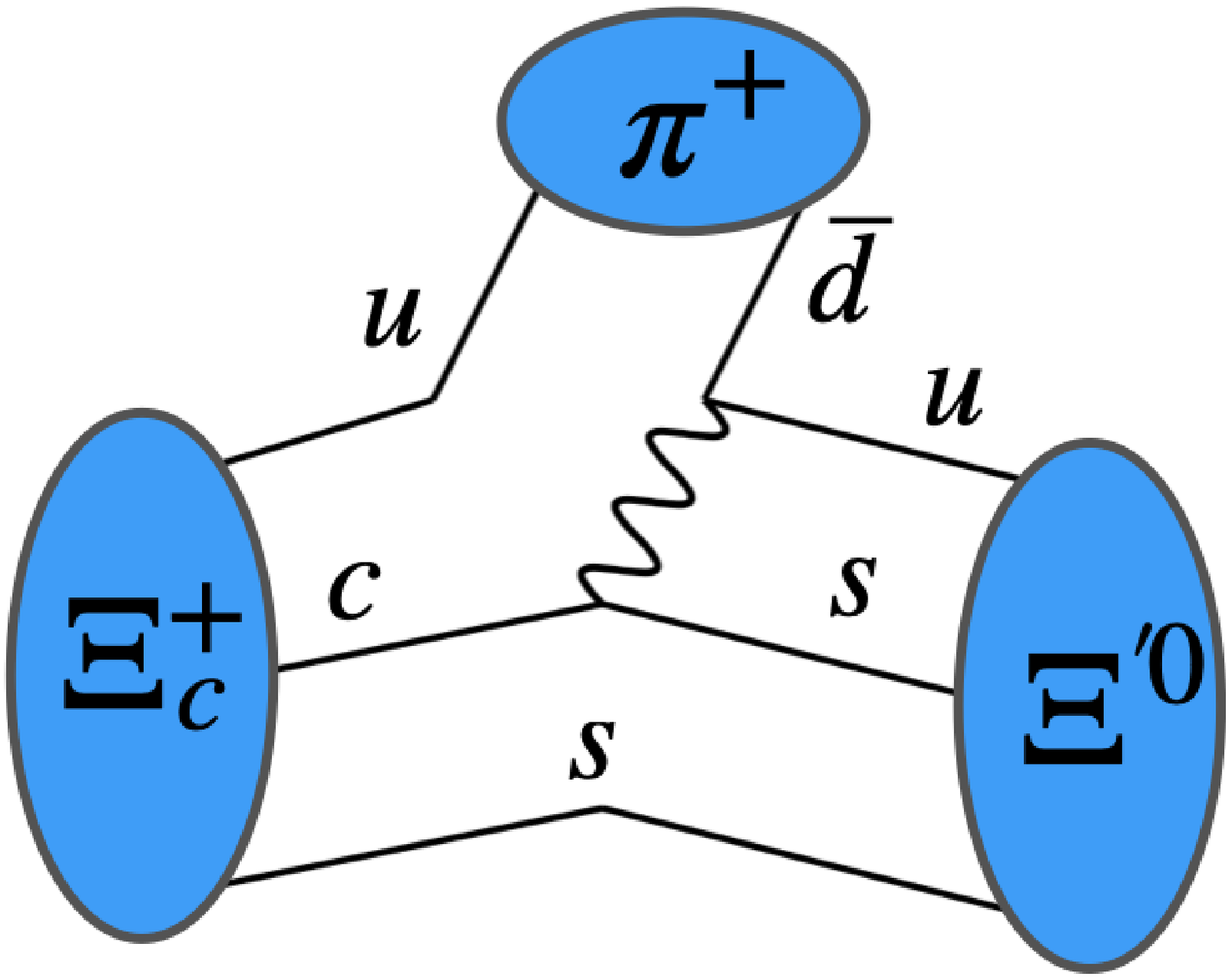}
	\caption{ The topological diagrams for $\Xi_c \to \Xi ' \ell ^+  \nu_\ell$
		and $\Xi_c \to \Xi' \pi^+$,  where   the blobs represent the hadronizations, and $\{ s, q \} $ indicates  that  $s$ and $q$ are symmetric in flavors and spins. 
	}
	\label{FIG3} 
\end{figure}

The topological diagram for $\Xi_c \to \Xi ' \ell^+ \nu_\ell$ is given in the left  hand side of FIG.~\ref{FIG3}.
Since the light quarks in $\Xi_c^{\overline 3}$ and $\Xi'$ are antisymmetric and symmetric  in flavors, respectively, the overlapping of the spectator quarks vanishes. 
We note that the gluon exchange corrections before and after the $W$ emission can be reabsorbed to the baryon wave functions and thus do not alter the argument. 
Accordingly, we have~\cite{SU(3)0}
\begin{equation}
\Gamma\left(\Xi_c^{\overline{{\bf 3}}} \to \Xi'  \ell^+ \nu_\ell   \right) = 0\,, 
\end{equation}
resulting in 
\begin{equation}
\Gamma\left(\Xi_c \to \Xi '  e^+ \nu_e \right)  =  \sin ^2 \theta_c \Gamma\left(\Xi_c^{{\bf 6}} \to \Xi '  e^+ \nu_e  \right) \,.
\end{equation}
With the similar formulas given in the previous section, 
we obtain that 
\begin{equation}
{\cal B} ( \Xi_c^0 \to \Xi ^{\prime -}  e^ + \nu_e  )  = (4.4\pm 0.5)  \times 10 ^{-3}\,,\qquad
{\cal B} ( \Xi_c^+ \to \Xi ^{\prime 0}  e^ + \nu  _e )  =  (1.3 \pm 0.2)\% \,,
\end{equation}
with $\theta_c = -0.137\pi$, and 
\begin{equation}
	{\cal B} ( \Xi_c^0 \to \Xi ^{\prime -}  e^ + \nu_e  )  = (8.7\pm 1.0)  \times 10 ^{-3}\,,\qquad
	{\cal B} ( \Xi_c^+ \to \Xi ^{\prime 0}  e^ + \nu  _e )  =  (2.6 \pm 0.4)\% \,,
\end{equation}
with $\theta_c = -0.2 \pi$. 
which are accessible at Belle~\MakeUppercase{\romannumeral 2}.  
We emphasize that  nonzero branching fractions of $\Xi_c \to \Xi' e^+ \nu_e$ in future experiments will be a smoking gun of $\theta_c \neq 0 $.

The nonleptonic decays of $\Xi_c^{+} \to \Xi^{\prime 0 }\pi^+ $ and $ \Xi_c^{+} \to \Sigma^{\prime +} \overline{K}^0$  are also forbidden for the absence of the mixing~\cite{Geng:2019awr}. To see it, the topological diagrams of $\Xi_c^{+} \to \Xi^{\prime 0 }\pi^+ $  are depicted in FIG.~\ref{FIG3}, where
the factorizable diagram is 
 obtained by replacing $\ell^+ \nu_\ell $ with $\pi^+~(u \overline{d})$, and  the right  figure is  the  nonfactorizable diagram.
 Due to the K\"orner-Pati-Woo theorem~\cite{Korner:1970xq}, the nonfactorizable amplitude vanishes, and the overlapping of the spectator quarks is zero in the left diagram  for $\Xi_c^{\overline{{\bf 3}}}$. 
Thus, we have 
\begin{equation}
	\begin{aligned}
&\Gamma \left( \Xi_c^{+} \to \Xi^{\prime 0 }  \pi^+
\right)  =\sin^2  \theta_c \Gamma \left( \Xi_c^{{\bf 6} +} \to \Xi^{\prime 0 } \pi ^+ 
\right) \,,\\
&\Gamma \left( \Xi_c^{+} \to \Sigma^{\prime + }\overline{K}^0 
\right)  =\sin^2  \theta_c \Gamma \left( \Xi_c^{{\bf 6} +} \to \Sigma^{\prime +} \overline{K}^0 
\right) \,,
	\end{aligned}
\end{equation} 
which can be calculated in the factorization framework. 

   The helicity amplitudes of 
$\Xi_c^{+} \to \Xi^{\prime 0 }  \pi^+$ and $ \Xi_c^{+} \to \Sigma^{\prime + }\overline{K}^0 $
 are then given  as 
\begin{eqnarray}\label{nonleptonic}
H_\pm = i \frac{G_F}{\sqrt{2}} V_{cs}a_1  f_\pi q^\mu \left \langle \Xi^{\prime 0 } , J_z = \pm \frac{1}{2}, \vec{p} = p \hat{z} \right |\overline{s} \gamma_\mu (1-\gamma_5) c \left | \Xi_c^{{\bf 6}+} , J_z = \pm \frac{1}{2} \right \rangle  \,,\nonumber\\
H_\pm = i \frac{G_F}{\sqrt{2}} V_{cs}a_2  f_K q^\mu \left \langle \Sigma^{\prime +} , J_z = \pm \frac{1}{2}, \vec{p} = p \hat{z} \right |\overline{u} \gamma_\mu (1-\gamma_5) c \left | \Xi_c^{{\bf 6} +} , J_z = \pm \frac{1}{2} \right \rangle  \,, 
\end{eqnarray}
respectively, 
 where $(a_{1}, a_2)= ( 1.26\pm 0.01 , 0.45\pm 0.03)$ are the effective Wilson coefficients~\cite{Cheng:2018hwl},  and $f_{\pi(K)}$ corresponds to the decay constant of the pion(kaon). 
The decay widths are given as 
\begin{equation}
\Gamma  = \frac{|\vec{p}| }{16 M_{\Xi_c^+}^2 \pi  }\left(  | H_+^2 | + |H_-^2|
\right) 
\,,
\end{equation}
leading to 
\begin{eqnarray}\label{nonle}
{\cal B} (\Xi_c^ + \to \Xi^{\prime 0} \pi ^+ ) = (3.8 \pm 0.5)\times 10^{-3}\,,~~~
{\cal B} (\Xi_c^ + \to \Sigma^{\prime +} \overline{K}^0 ) = (6.6 \pm 1.0)\times 10^{-4}\,.
\end{eqnarray}
with $|\theta_c| = 0.137 \pi$, and 
\begin{eqnarray}\label{nonle}
	{\cal B} (\Xi_c^ + \to \Xi^{\prime 0} \pi ^+ ) = (7.5 \pm 1.0)\times 10^{-3}\,,~~~
	{\cal B} (\Xi_c^ + \to \Sigma^{\prime +} \overline{K}^0 ) = (1.3 \pm 0.2)\times 10^{-3}\,,
\end{eqnarray}
with $|\theta_c| = 0.2 \pi$.

On the other hand,  in 2003 the FOCUS collaboration measured the ratios~\cite{FOCUS:2003gpe}
\begin{equation}\label{38}
\begin{aligned}
\frac{\Gamma(\Xi_c^+ \to \Xi^{'0} \pi^+ )}{\Gamma(\Xi_c^+ \to \Xi^{-} \pi ^+  \pi^+ )} < 0.1\,, ~~~
\frac{\Gamma(\Xi_c^+ \to \Sigma^{'+} \overline{K}^0)}{\Gamma(\Xi_c^+ \to \Xi^{-} \pi ^+  \pi^+ )} =1.00 \pm 0.49\,.  
\end{aligned}
\end{equation} 
Combining Eq.~\eqref{38} with the recent experimental absolute branching fraction~\cite{XPP} 
\begin{equation}
{\cal B}(\Xi_c^+ \to \Xi^- \pi^+\pi^+) = (2.9\pm 1.3)\%\,,
\end{equation}
we arrive at 
\begin{eqnarray}
{\cal B} (\Xi_c^+ \to \Xi^{\prime 0} \pi ^+ ) < 4.2 \times  10^{-3}\,,~~~
{\cal B} (\Xi_c^+ \to \Sigma^{\prime +} \overline{K}^0 ) = (2.9 \pm 2.0) \%\,.
\end{eqnarray}
The upper bound of ${\cal B} (\Xi_c^+ \to \Xi^{\prime 0} \pi ^+ )$ is very close to our result with $|\theta_c| = 0.137\pi$, but too small compare to the one with $|\theta_c |= 0.2 \pi$.  
We note in FIG.~2 of Ref.~\cite{newBelle}, the Belle collaboration has already detected nonzero signals of $\Xi_c^+ \to \Xi^{\prime 0} \pi^+$, 
but a definite number for its branching fraction is  still lacking.\footnote{
In FIG.~2 of Ref.~\cite{newBelle}, the dashed red curve is  contributed by  $\Xi_c^+ \to \Xi^{\prime0}  ( \to \Xi^- \pi^+) \pi ^+$.  At $M(\Xi^-\pi^+) = 1.53$~GeV, the figure shows that the number of the  observed events per $0.003$~GeV is roughly 400.} 
On the other hand the experimental value  of ${\cal B} (\Xi_c^+ \to \Sigma^{\prime +} \overline{K}^0 ) $ supports the existence of the mixing. However, the uncertainties are  too large at the current stage to draw a firm conclusion. 
We strongly recommend the future experiments to measure  $\Xi_c\to \Xi' e^+ \nu_e$ and pay a revisitation on $\Xi_c^+ \to \Xi' \pi^+ $ and  $\Xi_c^+ \to \Sigma^{\prime +} \overline{K}^0 $. 

By far,
the decay widths discussed in this section are in the second order of $\theta_c$. In some other decays, the baryonic mixing effects can be in the first order.
For instance, 
in $\Xi_c$ decays~\cite{Xic0,Xic1,Xic23,Xic3,Xic4,Xic5,Xic6}, 
the  effects are in the first order and could be responsible for the large $SU(3)_F$ breaking of ${\cal B}(\Xi_c^0 \to \Xi^- K^+)/{\cal B}(\Xi_c^0 \to \Xi^- \pi^+)$~\cite{breaking0,breaking1}. 
Furthermore, in  $\Xi_{cc} \to \Xi_c \pi$~\cite{Xicc0,Xicc1,Xicc2} the effects are also  in the first order of $\theta_c$ and can be responsible for the ratio of
${\cal B}(\Xi_{cc}^{++} \to \Xi_c^{\prime +} \pi^+)/{\cal B}(\Xi_{cc}^{++} \to \Xi_c^+ \pi^+)=1.41\pm 0.17\pm0.10$ measured at LHCb~\cite{LHCb:2022rpd,Ke:2022gxm}. Clearly, once the baryon mixing is confirmed by the future experiments, these decays should be paid revisitations by the theoretical studies.

\section{conclusion}

We have proposed the existence of the $\Xi_c-\Xi_c'$ mixing to resolve the tension between the experimental measurements and theoretical expectations in $\Xi_c^0 \to \Xi_c^- e^+ \nu_e$. 
We have analyzed the $\Xi_Q- \Xi_Q'$ mixings  from the mass spectra and obtained $|\theta_c |=  0.137(5)\pi$ and $|\theta_b | = 0.049(13)\pi$.    
With  $\theta_c = -0.137\pi$,
we  have found  that ${\cal B}(\Xi^0_c \to \Xi^- e^+ \nu_e) = (3.02 \pm 0.58) \%$ and ${\cal B}(\Xi^+_c \to \Xi^0 e^+ \nu_e ) = (9.00 \pm 1.74) \%$, which  are consistent with the results by the ALICE collaboration~\cite{ALICE:2021bli} and LQCD~\cite{Lattice}. 
We have shown that  $R(\theta_c = -0.137\pi )= 0.70\pm 0.09$ and $R(\theta_c =-0.2\pi) = 0.57\pm 0.08$, which  are  consistent with $R_{av}^{(\prime)}=0.59\pm 0.10~(0.46\pm 0.07)$  from the experiments~\cite{ALICE:2021bli,Belle:2021crz} and LQCD~\cite{Lattice},  and   successfully resolve the puzzle in $\Xi_c^0 \to \Xi ^- e^+ \nu_e$. 

We recommend the future experiments to measure $\Xi_c \to \Xi' e^+ \nu_e$, of which the branching fractions for   $\Xi_c^0$ and $\Xi_c^+$  have been evaluated to be $(4.4\pm 0.5) \times 10^{-3}$  and $(1.3\pm 0.2)\%$ with $|\theta_c| = 0.137\pi $, and $(8.7\pm1.0)\times 10^{-3}$ and $(2.6\pm0.4)\%$ with $|\theta_c| = 0.2 \pi$, respectively.
A nonvanishing branching fraction of these channels in experiments will be a smoking gun of the  $\Xi_c-\Xi_c'$ mixing.  In addition, the nonleptonic decays of $\Xi_c^+ \to \Xi' \pi^+$ and $\Xi_c^+ \to \Sigma^{\prime +} \overline{K}^0$ are forbidden without the mixing also.
We have shown that ${\cal B}(\Xi_c^+ \to \Xi^{\prime 0}\pi^+) = (3.8\pm 0.5)\times 10^{-3}$ and ${\cal B} (\Xi_c^ + \to \Sigma^{\prime +} \overline{K}^0 ) = (6.6 \pm 1.0 )\times 10^{-4}$ with  $|\theta_c| = 0.137\pi$, and 
${\cal B}(\Xi_c^+ \to \Xi^{\prime 0}\pi^+) = (7.5\pm 1.0)\times 10^{-3}$ and ${\cal B} (\Xi_c^ + \to \Sigma^{\prime +} \overline{K}^0 ) = (1.3 \pm 0.2 )\times 10^{-3}$ with $|\theta_c| = 0.2\pi$. 
These decay channels are accessible at Belle, Belle II, and LHCb. It is worth to note that the Belle collaboration has already observed signals of  $\Xi_c^+ \to  (\Xi^- \pi^+ \pi^+, \Lambda K_S^0 \pi^+)$~\cite{Belle:2016lhy}, which can come from the cascade decays of $\Xi_c^+ \to (\Xi^{\prime 0 } \pi^+, \Sigma^{\prime +} \overline{K}^0 )$, respectively.

We emphasize that the mixing is sizable and shall be considered seriously in the   studies of the charmed baryons.  In particular, most of the $SU(3)_F$ relations  in the  literature are broken~\cite{Xic0}, and revisitations are clearly  required. 
Finally, we remark that although the mixing of $\Xi_b-\Xi_b'$ is found to be small, its effects should be also explored in the future studies.

\begin{acknowledgments}
	We would like to thank Long-Ke Li  and Bing-Dong Wan for  valuable discussions.
	This work is supported in part by the National Key Research and Development Program of China under Grant No. 2020YFC2201501 and  the National Natural Science Foundation of China (NSFC) under Grant No. 12147103.
\end{acknowledgments}

\end{document}